\documentclass[aps,prl,twocolumn,amssymb, amsmath, secnumarabic, nobibnotes, superscriptaddress,groupedaddressl]{revtex4-2}  
\usepackage{graphicx}
\usepackage{dcolumn}   
\usepackage{bm}       
\usepackage{amssymb}

\usepackage{graphicx}
\usepackage[colorlinks,bookmarks=false,citecolor=blue,linkcolor=blue,urlcolor=black]{hyperref}

\DeclareGraphicsExtensions{.pdf}

\usepackage{xcolor}

\usepackage{xprintlen}
\begin{document}

\title{Polaronic Enhancement of Second-Harmonic Generation in Lithium Niobate}

\author{Agnieszka L. Kozub}
\email{agnieszka.kozub@upb.de} 
\affiliation{%
	Universit\"at Paderborn, Department Physik, 33095 Paderborn, Germany
}%

\author{Arno Schindlmayr}
\affiliation{%
	Universit\"at Paderborn, Department Physik, 33095 Paderborn, Germany
}%

\author{Uwe Gerstmann}%
\affiliation{%
	Universit\"at Paderborn, Department Physik, 33095 Paderborn, Germany 
}%

\author{Wolf Gero Schmidt}%
\affiliation{%
	Universit\"at Paderborn, Department Physik, 33095 Paderborn, Germany 
}%
                     
\date{\today}

\begin{abstract}
Density-functional theory within a Berry-phase formulation of the dynamical polarization
is used to determine the second-order susceptibility $\chi^{(2)}$ of lithium niobate (LiNbO$_3$).
Defect trapped polarons and bipolarons are found to strongly enhance the nonlinear susceptibility
of the material, in particular if localized at Nb$_\mathrm{V}$--V$_{\mathrm{Li}}$ defect pairs.
This is essentially a consequence of the polaronic excitation resulting in relaxation-induced 
gap states. The occupation of these levels leads to strongly enhanced $\chi^{(2)}$ coefficients 
and allows for the spatial and transient modification of the second-harmonic generation of
macroscopic samples.                
\end{abstract}

\maketitle

The arbitrary control of 
electromagnetic waves is a key aim of photonic research.
This requires the engineering of 
optical coefficients, e.g., by doping, strain, or 
electric fields \cite{Cazzanelli2012,Jacobsen2006,Seyler2015,Runge2019}.
Photorefractive materials like lithium niobate (LiNbO$_3$, LN) that respond to light 
by altering their refractive index are of particular interest in this context \cite{PR06,PR14}.
Lithium niobate is used, e.g., for optical waveguides, optical modulators,
photonic integrated circuits and nonvolatile holographic storage \cite{Hesselink_1998,Buse_1998}.

\begin{figure}[thb]
	\centering
	\includegraphics[width=8.6cm]{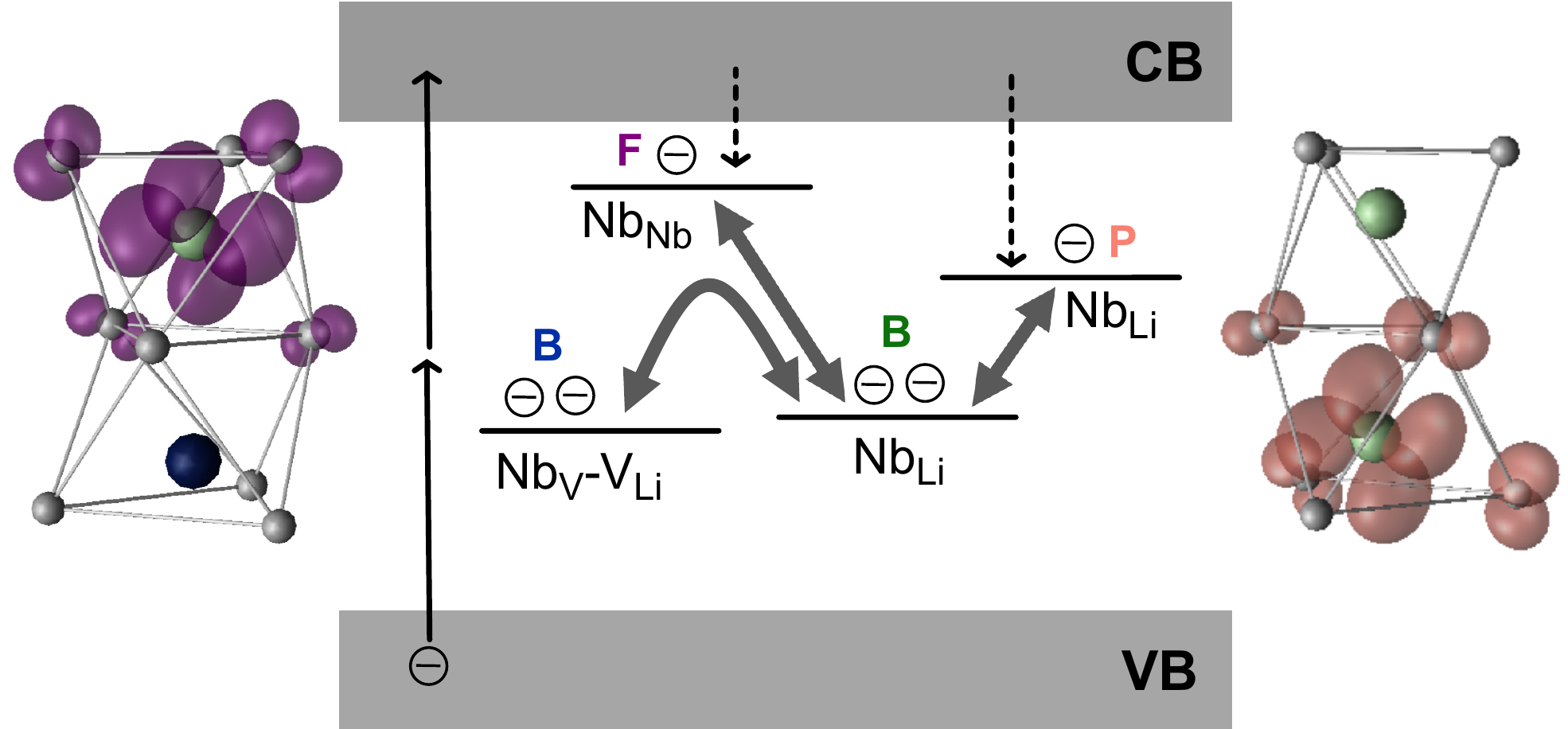}
	\caption{\label{bandschema} (Color online) Schematic illustration of electron polaron states in lithium niobate. Band-to-band
	excitation and subsequent carrier localization result in the formation of free (F) and 
	bound (P) polarons, which combine to bipolarons (B). Bipolarons may transform structurally or dissociate into
	single electron polarons. Charge densities of the polaron orbitals are shown on the sides: free polaron (left) and bound polaron (right), both in tilted ground state configuration..
	See Fig. \ref{structures} for complete structures and the notation of atoms. } 
\end{figure}

The LN optical properties are profoundly affected by polarons, i.e.,    
electrons excited to a state close to the edge of the conduction band that get dressed 
with a cloud of virtual polar phonons \cite{Schirmer:2009iy,Buse_2012,Imlau-2015,Krampf:2020ip,Furukawa1992}. 
Free small polarons (F) are the simplest polaron species in LN, see Fig.\ \ref{bandschema}. They are formed by extra electrons stabilized at regular Nb$_{\mathrm{Nb}}$ ions of the LN lattice. Bound polarons (P) and bound bipolarons (B) are formed by single electrons and pairs of electrons, respectively, localized at point defects. These defects are related to excess Nb 
compensating the lithium deficit observed in the congruently melting composition of lithium niobate.
Single polarons (F and P) are usually found as metastable states upon optical excitation and may 
combine to bipolarons. The latter also result from thermal or electrochemical reduction 
and can be thermally or optically split into single 
electron polarons \cite{Schirmer:2009iy,Imlau-2015,PhysRevLett.96.186404}.
Optical absorption peaks at 
about 0.9 eV, 1.6 eV, and 2.5 eV are assigned to F \cite{Schirmer-1994}, P 
\cite{Halliburton1983}, and B polarons\cite{Koppitz_1987}, respectively.

Controlling second-harmonic generation (SHG) in lithium niobate, i.e., the 
generation of new photons with twice the energy of the initial photons,
requires engineering the second-order susceptibility $\chi^{(2)}$ \cite{Boyd,Sipe2000,Mendoza2019}.
Although polarons have been known for a long time to cause various 
optical nonlinearities, such as green-induced infrared absorption \cite{Furukawa_2001},
their influence on $\chi^{(2)}$ is essentially unknown.
In this Letter, we show by means of 
first-princples calculations that electron polarons cause a strong
enhancement of the $\chi^{(2)}$ susceptibility for photon energies in the lower half of the LN band gap
and thus provide a means to control SHG in space and time. 
 
In detail, we perform density-functional theory (DFT) electronic-structure 
calculations based on the Quantum Espresso \cite{Giannozzi_2009_short,Giannozzi_2017}
implementation. The generalized gradient approximation is employed using the PBEsol functional.
The nonlinear susceptibilities are obtained from the real-time evolution 
of the Bloch electrons in a uniform time-dependent electric field following the 
Berry-phase approach proposed by Attaccalite and Gr\"uning \cite{Attacalite}.

\begin{figure}[thb]
	\centering
	\includegraphics[width=8.6cm]{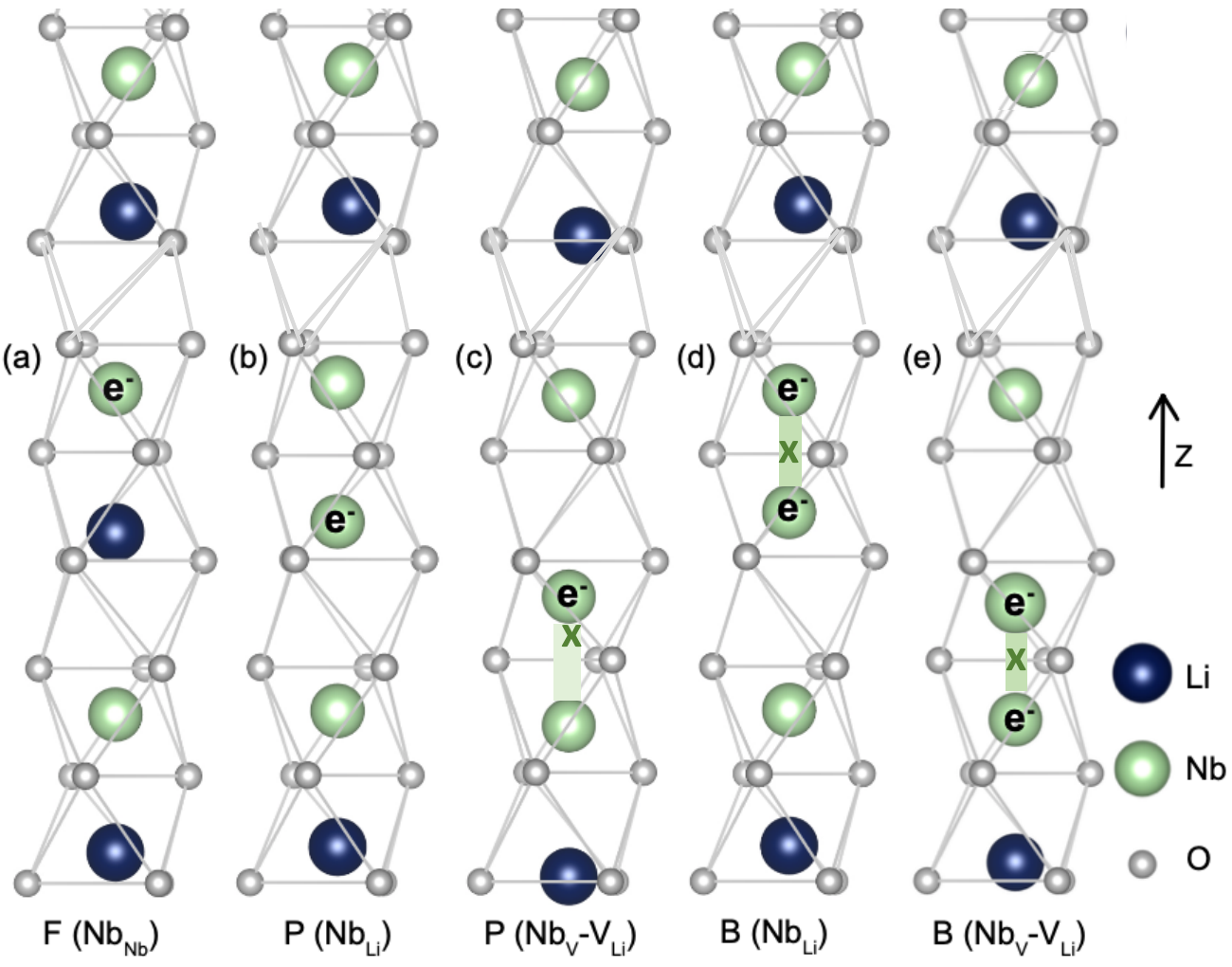}
	\caption{\label{structures} (Color online) Free polarons (a), bound polarons (b, c), and bipolarons (d, e) in LiNbO$_3$
          defect structures. Trapped electrons (e$^{-}$) as well as their hybridization along the $z$-axis and the center of the
          polaronic charges (x) are indicated (cf.~Fig.~\ref{SHG}). } 
\end{figure}

Stoichiometric LN crystallizes in a rhombohedral unit cell. Oxygen atoms form octahedra that are 
alternatingly occupied by Li, Nb, or are empty, see Fig.\ \ref{structures}(a). 
The present polaron calculations are based on previously established models \cite{Falko},
which account for both the measured optical absorption peaks and the electron paramagnetic resonance data.
As shown in  Fig.\ \ref{structures}, we consider
(a) free polarons, where the excess electron is bound to a regular Nb$_{\mathrm{Nb}}$ ion, 
(b) bound polarons with the excess electron at the Nb$_{\mathrm{Li}}$ antisite defect,
(c) bound polarons with the excess electron at the Nb interstitial of the Nb$_\mathrm{V}$--V$_{\mathrm{Li}}$ defect pair and bipolarons, where one of the excess electrons is at the defect niobium atom of
(d) the Nb$_{\mathrm{Li}}$ antisite defect or 
(e) the Nb$_\mathrm{V}$--V$_{\mathrm{Li}}$ defect pair, and the second one is at the neighboring Nb atom. The structures shown in (a),  (b), and (c) are studied in both the axially symmetric 
and quasi-Jahn-Teller distorted (tilted) configurations~\cite{Falko}. The structures (d) and (e) have axial symmetry. 
In this case, the niobium atoms carrying the excess electrons shift along the crystal $z$ axis preserving 
the threefold rotational symmetry. In the tilted configurations, the polaron-carrying atoms are shifted 
in the $xy$ plane, which results in clover-leave like orbitals for the excess electrons~\cite{Falko}.
The defect structures are modeled with 80-atom supercells, as described in Ref.~\cite{Falko_crys}. 
Here, a 3$\times$3$\times$3 $\Gamma$-centred $\mathbf{k}$-point sampling is used for the $\chi^{(2)}$ calculations.  
The bulk LN calculations are performed using the 10-atom unit cell 
and a 6$\times$6$\times$6 $\Gamma$-centred $\mathbf{k}$-point grid to determine $\chi^{(2)}$. 

\begin{figure}[ttt]
	\centering
	\includegraphics[width=8.6cm]{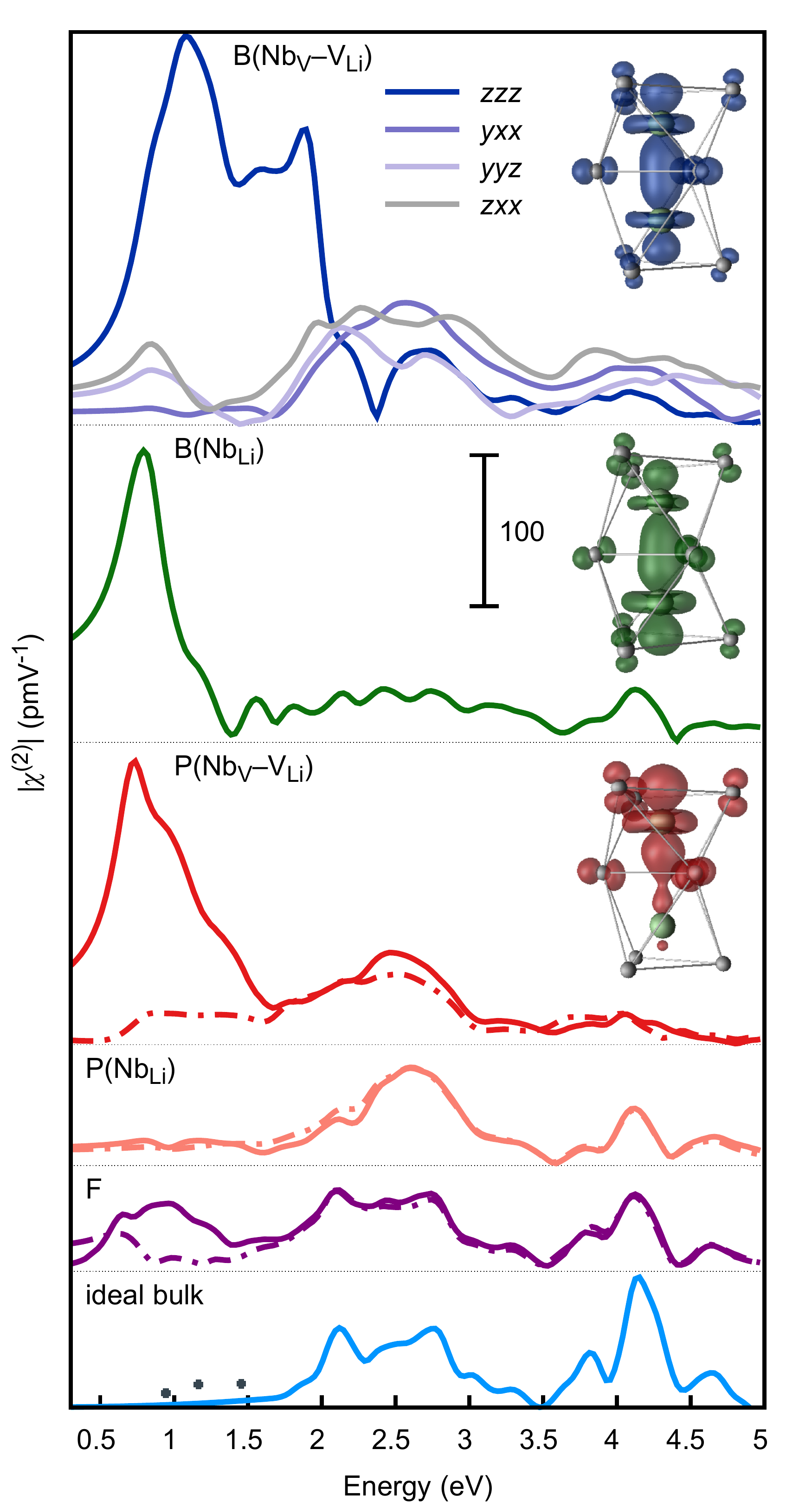}
	\caption{\label{SHG} (Color online) Values of $|\chi^{(2)}(\omega) |$ calculated for ideal, stoichiometric LN as well as for supercells that contain electron polarons. Solid lines correspond to the structures with the axial symmetry, dashed lines to the tilted configuration. Black symbols in the bottom represent experimental data for congruent LiNbO$_3$\cite{Shoji}.
	} 
\end{figure}

Fig.\ \ref{SHG} shows the modulus of calculated second-order susceptibility tensor elements in dependence on the incident photon energy. 
Unless stated otherwise, we plot the most frequently measured value $|\chi^{(2)}_{zzz} (\omega) |$.  The calculations  for the
stoichiometric bulk material agree well with previous calculations \cite{Riefer_2013} and reproduce the order of magnitude 
of the measured data~\cite{Shoji}.  
The low $|\chi^{(2)}_{zzz} (\omega) |$ values for energies in the lower half of the LN band gap are strongly enhanced upon bipolaron
formation and for P(Nb$_\mathrm{V}$--V$_{\mathrm{Li}}$) with the axial symmetry. 
In the case of B(Nb$_\mathrm{V}$--V$_{\mathrm{Li}}$), where the enhancement is most pronounced, we also show the other
second-order susceptibility tensor elements that are nonzero in trigonal symmetry~\cite{SHG_symmetry_1999},
as e.g. present for stoichiometric LN as well as for the polaron states with axial symmetry:
$|\chi^{(2)}_{yxx} (\omega) |$, $|\chi^{(2)}_{yyz} (\omega) |$,  and $|\chi^{(2)}_{zxx} (\omega) |$ in Fig.\ \ref{SHG}. 
The latter are far smaller than $|\chi^{(2)}_{zzz} (\omega) |$ in the low-energy region.  

A common feature of all the systems with strongly enhanced $|\chi^{(2)}_{zzz} (\omega) |$ is a pronounced hybridization of the polaron orbital
at the defect niobium atom with the neighbouring atom along the $z$ direction (see insets in Fig.\ \ref{SHG}). This hybridization shifts the
center of the polaronic charge away from its position in ideal stochiometric LN. 

\begin{figure}[tb]
	\centering
	\includegraphics[width=8.6cm]{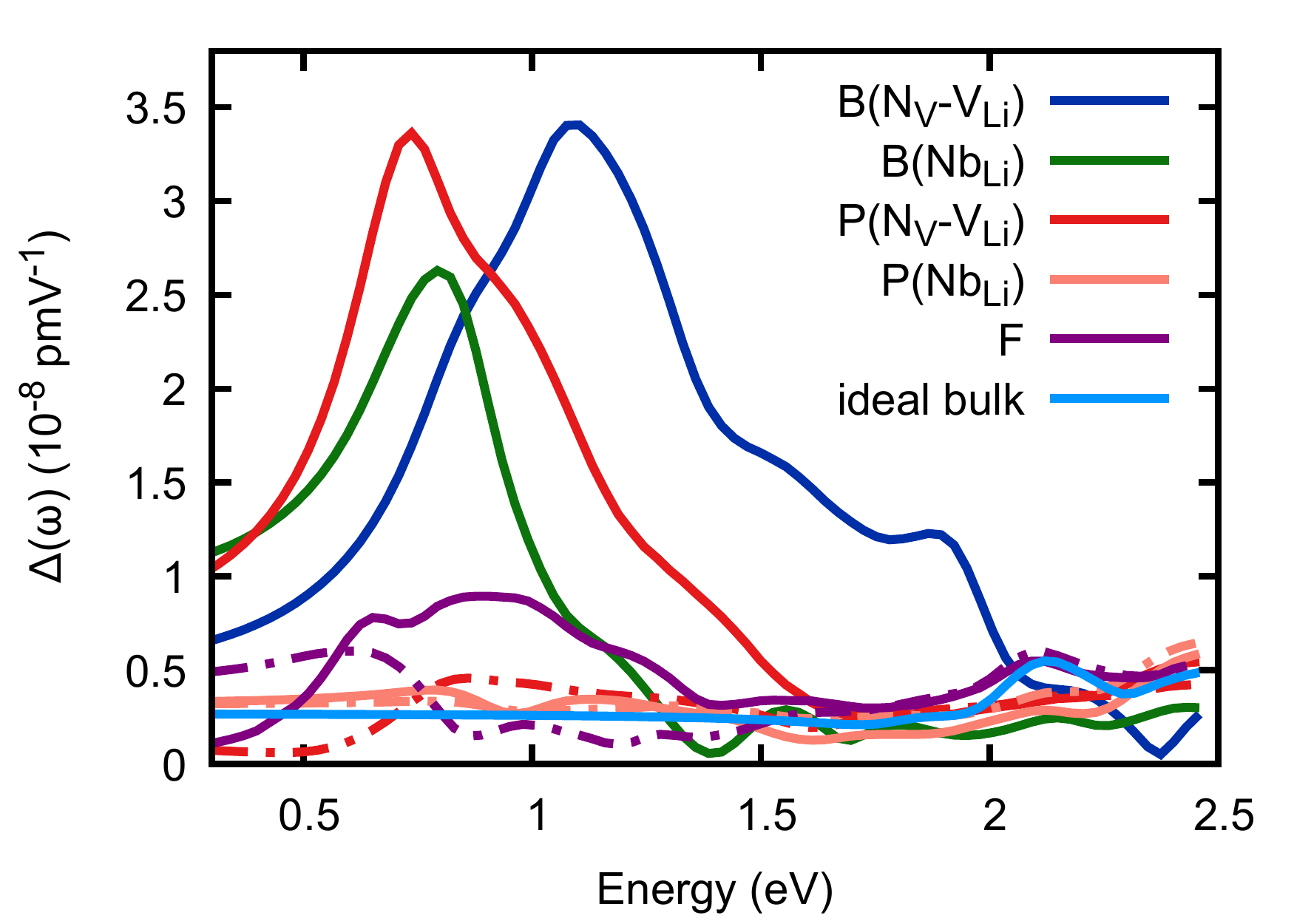}
	\caption{\label{Miller} (Color online) Check of Miller’s rule for bulk LiNbO$_3$ as well as 
	crystal structures containing electron polarons.
	} 
\end{figure}

According to Miller's rule \cite{Miller_1964,Boyd}, the quantity 
$\Delta(\omega) = \chi^{(2)}(2\omega)/\{ \chi^{(1)}(2\omega)[\chi^{(1)}(\omega)]^2 \}$
is a slowly varying function of $\omega$. It accounts for frequency-doubling and allows in many cases to relate linear
and  nonlinear susceptibilities of noncentrosymmetric crystals. As can be seen in Fig.\ \ref{Miller},
Miller's rule holds indeed for stoichiometric LN but is clearly violated for 
bipolarons and Nb$_{\mathrm{V}}$--V$_{\mathrm{Li}}$  localized bound polarons.
Obviously, these systems are characterized by a strong bond charge acentricity and are not well 
described within the classical anharmonic oscillator model \cite{Levine1973,Boyd}. 
In fact, the $\chi^{(2)}$ values calculated here for the polaron or bipolaron containing supercells, i.e., 
$|\chi^{(2)}_{zzz} | \approx 200$~pmV$^{-1}$ within the region of the LN band gap, are 
comparable to the value of GaAs in the energy region above the 
band gap, which is known as one of the largest $\chi^{(2)}$ values reported for solid state systems~\cite{Boyd}.

\begin{figure}[bbb]
	\centering
	\includegraphics[width=8.6cm]{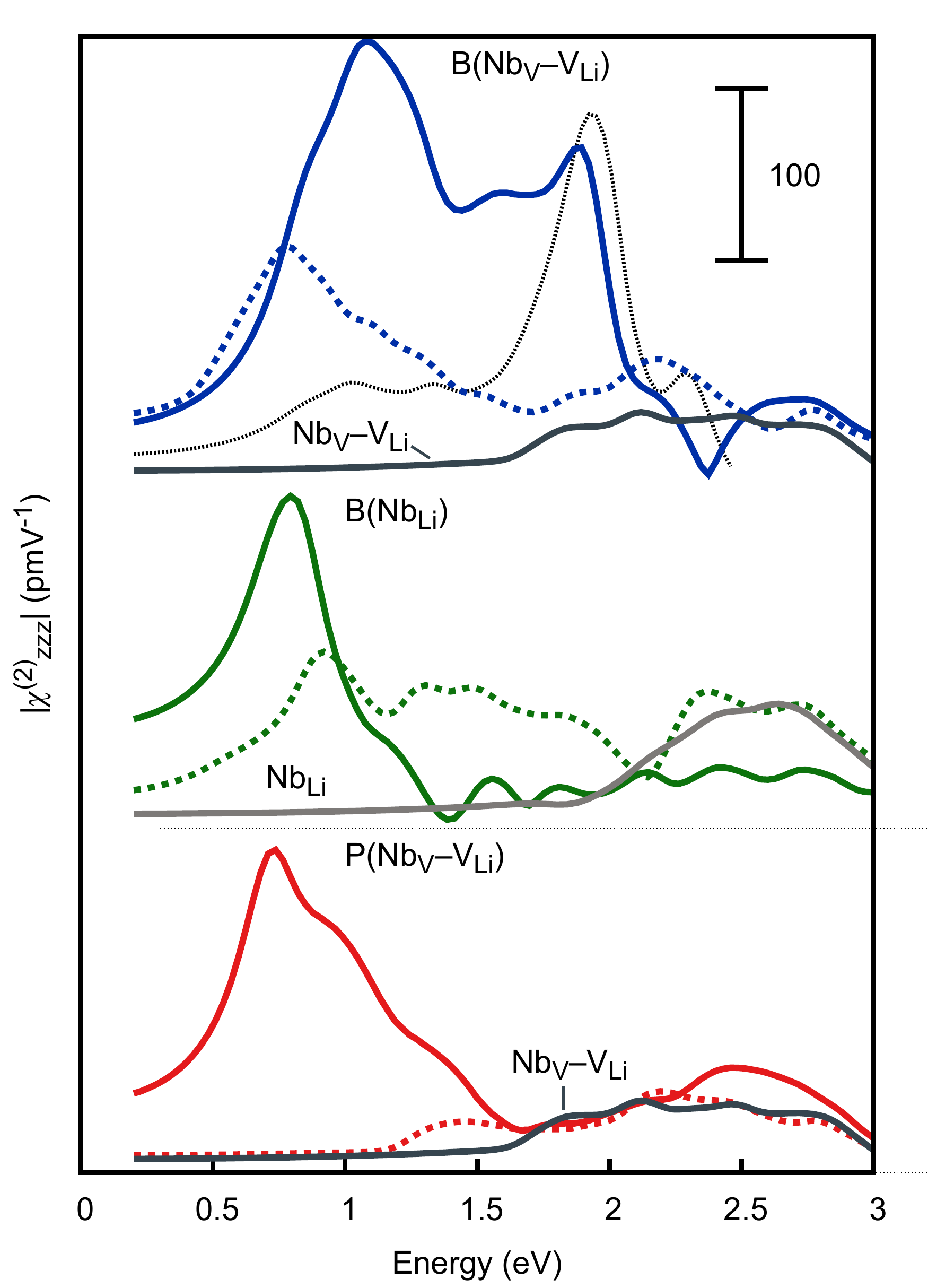}
        \vspace*{-4mm}
        \caption{\label{step_by_step} (Color online) $|\chi^{(2)}_{zzz}(\omega) |$ calculated for supercells 
	containing defects, defect localized (bi)polarons, and (shown with colored dashed lines) polaronic structures without trapped electrons.
	The black dashed line represents $|\chi^{(2)}_{zzz}(\omega) |$ estimated from Miller's rule for B(Nb$_{\mathrm{V}}$--V$_{\mathrm{Li}}$).}
	\vspace*{-2mm} 
\end{figure}

{\em What is the microscopic origin  of the large optical nonlinearities associated with some polaron species?}
In order to answer this question we separate the influence of the defect formation and of the 
polaronic excitation, i.e., the trapping of additional electrons and of the associated lattice 
relaxation for the cases of the bipolarons as well as 
P(Nb$_{\mathrm{V}}$--V$_{\mathrm{Li}}$). In Fig.\ \ref{step_by_step} it can be seen that the point defects themselves 
only slightly modify the $|\chi^{(2)}_{zzz}|$ bulk values, at least for low photon energies.
However, upon polaronic excitation the low-energy susceptibility is much enhanced, in analogy to the 
polaron-related additional absorption peaks 
in the linear optical response \cite{Schirmer-1994,Halliburton1983,Koppitz_1987}. 
This enhancement can either be due to electronic transitions related to the polaron states, or it may be
caused by the lattice relaxation accompanying the polaron formation. To discriminate between these
two possibilities we perform calculations for supercells that model the (bi)polaron geometries, but do not
contain the corresponding electrons. The corresponding spectra are plotted with dashed lines in 
Fig.\ \ref{step_by_step}. It can be seen that in the case of bipolarons both electronic occupation
and lattice relaxation contribute to the $\chi^{(2)}$ enhancement, while the electronic effects 
are dominant for bound polarons. 
For B(Nb$_{\mathrm{V}}$--V$_{\mathrm{Li}}$) the SHG enhancement is not restricted to the spectral region around 1 eV,
but extends to energies up to 2 eV, where an additonal second peak can be identified. This 
peak is related to transitions into empty Nb 4$d$ states close the conduction-band edge and be 
already identified in the B(Nb$_{\mathrm{V}}$--V$_{\mathrm{Li}}$) linear optical response \cite{Falko_crys}.
Indeed, it well predicted already from Miller's rule applied to $\chi^{(1)}$ calculated for B(Nb$_{\mathrm{V}}$--V$_{\mathrm{Li}}$).
This is shown by the black dashed line in Fig. \ref{step_by_step}.

\begin{figure}[htb]
	\centering
	\includegraphics[width=8.6cm]{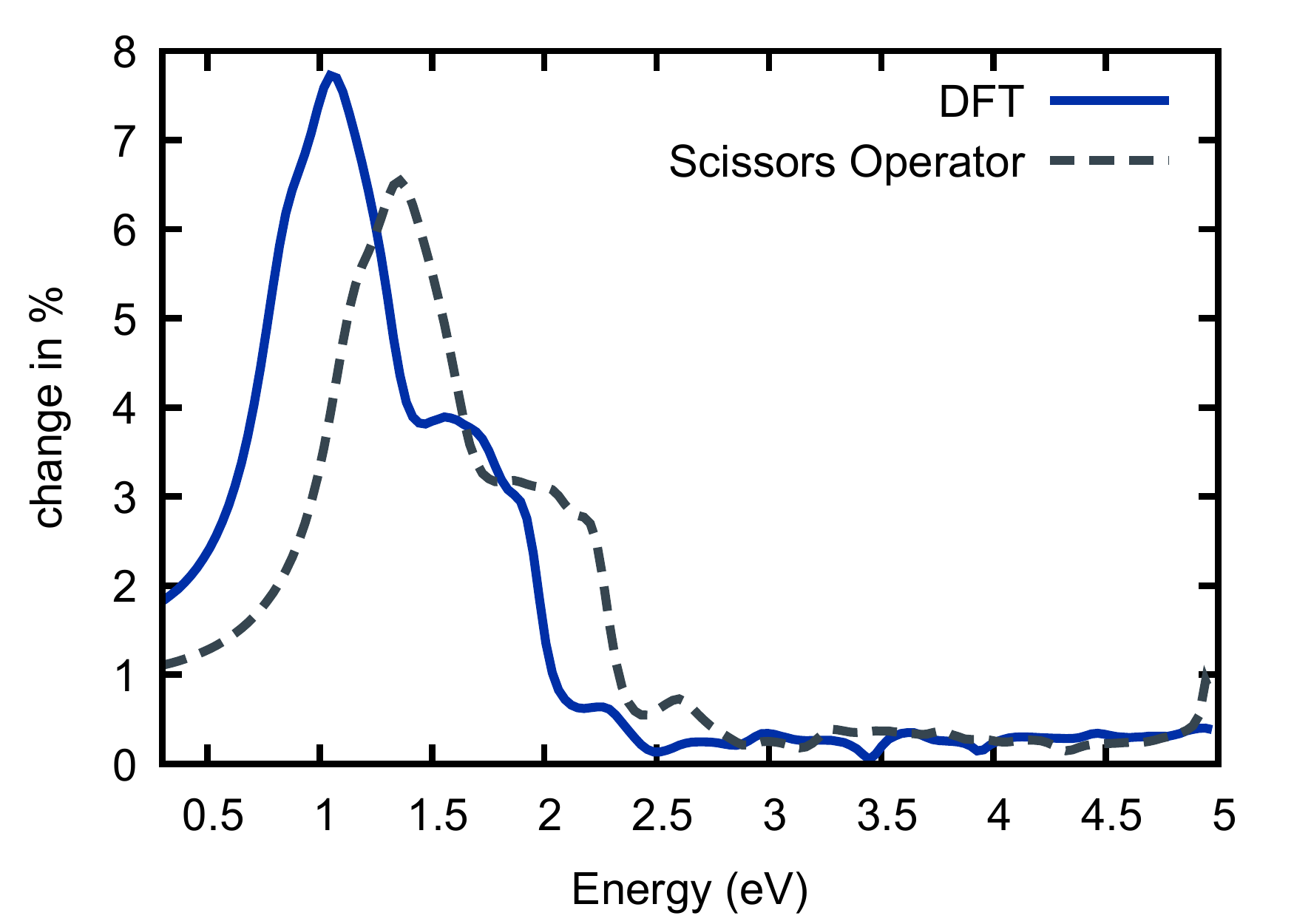}
	\caption{\label{change} (Color online) Relative  change in $|\chi^{(2)}_{zzz} (\omega) |$ 
	due to B(Nb$_{\mathrm{V}}$--V$_{\mathrm{Li}}$) formation in bulk LiNbO$_3$, assuming a density of 4.4$\times 10^{24}$ m$^{-3}$. 
	In addition to calculations within the independent-particle approximation (DFT) also an estimate for the influence of 
	self-energy corrections (Scissors Operator) is shown, see text. 
	} 
\end{figure}

{\em How important will the polaronic SHG enhancement effect be for real samples?}
Merschjann and co-workers~\cite{Merschjann_2008}
determined polaron densities in congruently melting LiNbO$_3$ by time-resolved pump–multiprobe 
spectroscopy. They obtained a steady-state number density of bipolarons of 6.7$\times 10^{23}$ m$^{-3}$ 
at room temperature without illumination.
The polaron density can be drastically enhanced by optical excitation.  
Electron polarons in lithium niobate are generated optically within a few hundred 
femtoseconds \cite{Beyer:2006dca,Badorreck}. Thereby concentrations
far higher than in the steady state can be achieved. Buse {\em et al.} \cite{Beyer:2006dca}, for example,
determine a polaron density of 4.4$\times 10^{24}$ m$^{-3}$. 
This value is used for an order-of-magnitude estimate for the bipolaron induced $|\chi^{(2)}_{zzz} (\omega) |$ change in Fig.\ \ref{change}. 
Obviously, $|\chi^{(2)}_{zzz}|$ changes of the order of a few percent can be expected to result
from optical excitation of real samples.  

A word of caution is in order considering the present DFT calculation. Optical excitations are known
to be influenced by electronic many-body effects beyond DFT  \cite{PhysRevB.65.035205,PhysRevB.71.195209,PhysRevB.89.235410,Riefer_2017,DEAK2018,Furth2014,Lambrecht2018}. 
In order
to probe their influence, a scissors operator correction
is added to the effective Hamiltonian for the Bloch electrons. Its size has been 
determined by $GW$ calculations \cite{Falko} that predict a 
blueshift of 0.6 eV for transitions from the B(Nb$_{\mathrm{V}}$--V$_{\mathrm{Li}}$) polaron
state to the LN conduction band minimum compared to the DFT calculations. 
As expected, the self-energy effects lead to a blueshift and weakening of the SHG signal.  
As excitonic effects are 
expected to redshift and increase the 
SHG spectral features \cite{PhysRevB.65.035205,PhysRevB.71.195209,PhysRevB.89.235410,Riefer_2017},
the dashed line in Fig.\ \ref{change} represents an upper limit for the 
excitation energies and a lower limit for the SHG enhancement that can be expected to result 
from the bipolaron formation. Obviously, the polaronic effect on $\chi^{(2)}$ predicted 
here from DFT is robust with respect to many-body effects. 

In summary, the present density-functional calculations predict a strong $\chi^{(2)}$ enhancement
in lithium niobate upon formation of bound (bi)polarons. It is caused primarily by
the occupation of the relaxation-induced polaronic defect levels in the gap.
Further details of the SHG enhancement depend sensitively on the relative position of the
polaronic charge density with respect to the neighboring cations of the ferroelectric host material.
The polaron density in real samples depends (i) on the
availability of possible lattice trapping sites -- that may be locally modified by doping \cite{Schirmer:2009iy} --
and (ii) can be strongly increased by optical excitation of the sample.
The polaronic enhancement of the nonlinear optical coefficients predicted here thus suggests 
a novel way to control second-harmonic generation in space and time beyond 
strain or external electric field induced effects.\\

The DFG (TRR 142, project number 231447078)
is acknowledged for financial support. 
We thank the Paderborn Center for Parallel Computing (PC$^2$) and the H\"ochstleistungs-Rechenzentrum Stuttgart (HLRS) 
for grants of high-performance computer time.

\bibliographystyle{apsrev4-2}
\bibliography{literature}
\end{document}